\documentclass[12pt,a4paper]{article}
\usepackage{amsmath}
\usepackage{xspace}
\usepackage{epsfig}
\usepackage{rotating}
\usepackage{dcolumn}
\usepackage{url}
\usepackage{hyperref}
\hypersetup{ pdfborder=0 0 0, urlbordercolor=1 0 0 }
\usepackage{cite}

\newcommand{\etal}{{\it et al.}}
\newcommand{\munu}{\mu^-\overline{\nu}}
\newcommand{\Bs}{\overline{B}_s^0}

\usepackage{graphicx}
\usepackage{color}
\usepackage{colortbl}
\usepackage{afterpage}

\begin{document}


\begin{titlepage}
\belowpdfbookmark{Title page}{title}

\pagenumbering{roman}
\vspace*{-1.5cm}
\centerline{\large EUROPEAN ORGANIZATION FOR NUCLEAR RESEARCH (CERN)}
\vspace*{1.5cm}
\hspace*{-5mm}\begin{tabular*}{16cm}{lc@{\extracolsep{\fill}}r}
\vspace*{-12mm}\mbox{\!\!\!\epsfig{figure=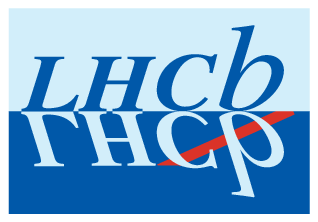,width=.12\textwidth}}& & \\
&& CERN-PH-EP-2011-008\\
&& 1 February 2011 \\
\end{tabular*}
\vspace*{3cm}
\begin{center}
{\bf\huge\boldmath {First observation of $\Bs\to D_{s2}^{*+} X\munu$ decays}\\
}
\vspace*{2cm}
\normalsize {
The LHCb Collaboration\footnote{Authors are listed on the following pages.}
}

\end{center}
\vspace{\fill}
\centerline{\bf Abstract}
\vspace*{5mm}\noindent
Using data collected with the LHCb detector in proton-proton collisions at a centre-of-mass energy of 7 TeV, the semileptonic decays  $\Bs\to D_s^+ X \munu$ and $\Bs\to D^0K^+ X \munu$ are detected. Two structures are observed in the $D^0K^+$ mass spectrum at masses consistent with the known $D_{s1}(2536)^{+}$ and $D^{*}_{s2}(2573)^+$ mesons. The measured branching fractions relative to the total $\Bs$ semileptonic rate are
${{\cal{B}}(\Bs\to D_{s2}^{*+} X \munu)}/{{\cal{B}}(\Bs\to X \munu)}= (3.3\pm 1.0\pm 0.4 )\%$, and
${{\cal{B}}(\Bs\to D_{s1}^+ X \munu)}/{{\cal{B}}(\Bs\to X \munu)}= (5.4\pm 1.2\pm 0.5)\%$,
where the first uncertainty is statistical and the second is systematic.
This is the first observation of the $D_{s2}^{*+}$ state in $\Bs$ decays; we also measure its
mass and width.

\vspace*{1.cm}
\noindent{\it Keywords:} LHC, semileptonic $b$ decays,  $\Bs$ meson \\
{\it PACS:} 14.40.Lb,  14.65.Fy,  13.20-He\

\vspace*{2.cm}
\center{To be published in Physics Letters B}
\vspace{\fill}

\end{titlepage}

\setcounter{page}{2}

\belowpdfbookmark{LHCb author list}{authors}
\centerline{\Large The LHCb Collaboration}
\begin{flushleft}
R.~Aaij$^{23}$, 
B.~Adeva$^{36}$, 
M.~Adinolfi$^{42}$, 
C.~Adrover$^{6}$, 
A.~Affolder$^{48}$, 
M.~Agari$^{10}$, 
Z.~Ajaltouni$^{5}$, 
J.~Albrecht$^{37}$, 
F.~Alessio$^{6,37}$, 
M.~Alexander$^{47}$, 
P.~Alvarez~Cartelle$^{36}$, 
A.A.~Alves~Jr$^{22}$, 
S.~Amato$^{2}$, 
Y.~Amhis$^{38}$, 
J.~Amoraal$^{23}$, 
J.~Anderson$^{39}$, 
R.~Antunes~Nobrega$^{22,l}$, 
R.B.~Appleby$^{50}$, 
O.~Aquines~Gutierrez$^{10}$, 
A.~Arefyev$^{30}$, 
L.~Arrabito$^{53}$, 
M.~Artuso$^{52}$, 
E.~Aslanides$^{6}$, 
G.~Auriemma$^{22,m}$, 
S.~Bachmann$^{11}$, 
D.S.~Bailey$^{50}$, 
V.~Balagura$^{30,37}$, 
W.~Baldini$^{16}$, 
R.J.~Barlow$^{50}$, 
C.~Barschel$^{37}$, 
S.~Barsuk$^{7}$, 
S.~Basiladze$^{31}$, 
A.~Bates$^{47}$, 
C.~Bauer$^{10}$, 
Th.~Bauer$^{23}$, 
A.~Bay$^{38}$, 
I.~Bediaga$^{1}$, 
K.~Belous$^{34}$, 
I.~Belyaev$^{30,37}$, 
M.~Benayoun$^{8}$, 
G.~Bencivenni$^{18}$, 
R.~Bernet$^{39}$, 
M.-O.~Bettler$^{17,37}$, 
M.~van~Beuzekom$^{23}$, 
S.~Bifani$^{12}$, 
A.~Bizzeti$^{17,h}$, 
P.M.~Bj\o rnstad$^{50}$, 
T.~Blake$^{49}$, 
F.~Blanc$^{38}$, 
C.~Blanks$^{49}$, 
J.~Blouw$^{11}$, 
S.~Blusk$^{52}$, 
A.~Bobrov$^{33}$, 
V.~Bocci$^{22}$, 
B.~Bochin$^{29}$, 
A.~Bondar$^{33}$, 
N.~Bondar$^{29,37}$, 
W.~Bonivento$^{15}$, 
S.~Borghi$^{47}$, 
A.~Borgia$^{52}$, 
E.~Bos$^{23}$, 
T.J.V.~Bowcock$^{48}$, 
C.~Bozzi$^{16}$, 
T.~Brambach$^{9}$, 
J.~van~den~Brand$^{24}$, 
J.~Bressieux$^{38}$, 
S.~Brisbane$^{51}$, 
M.~Britsch$^{10}$, 
T.~Britton$^{52}$, 
N.H.~Brook$^{42}$, 
H.~Brown$^{48}$, 
A.~B\"{u}chler-Germann$^{39}$, 
A.~Bursche$^{39}$, 
J.~Buytaert$^{37}$, 
S.~Cadeddu$^{15}$, 
J.M.~Caicedo~Carvajal$^{37}$, 
O.~Callot$^{7}$, 
M.~Calvi$^{20,j}$, 
M.~Calvo~Gomez$^{35,n}$, 
A.~Camboni$^{35}$, 
W.~Cameron$^{49}$, 
L.~Camilleri$^{37}$, 
P.~Campana$^{18}$, 
A.~Carbone$^{14}$, 
G.~Carboni$^{21,k}$, 
R.~Cardinale$^{19,i}$, 
A.~Cardini$^{15}$, 
L.~Carson$^{36}$, 
K.~Carvalho~Akiba$^{23}$, 
G.~Casse$^{48}$, 
M.~Cattaneo$^{37}$, 
M.~Charles$^{51}$, 
Ph.~Charpentier$^{37}$, 
J.~Cheng$^{3}$, 
N.~Chiapolini$^{39}$, 
A.~Chlopik$^{27}$, 
J.~Christiansen$^{37}$, 
P.~Ciambrone$^{18}$, 
X.~Cid~Vidal$^{36}$, 
P.J.~Clark$^{46}$, 
P.E.L.~Clarke$^{46}$, 
M.~Clemencic$^{37}$, 
H.V.~Cliff$^{43}$, 
J.~Closier$^{37}$, 
C.~Coca$^{28}$, 
V.~Coco$^{23}$, 
J.~Cogan$^{6}$, 
P.~Collins$^{37}$, 
F.~Constantin$^{28}$, 
G.~Conti$^{38}$, 
A.~Contu$^{51}$, 
M.~Coombes$^{42}$, 
G.~Corti$^{37}$, 
G.A.~Cowan$^{38}$, 
R.~Currie$^{46}$, 
B.~D'Almagne$^{7}$, 
C.~D'Ambrosio$^{37}$, 
I.~D'Antone$^{14}$, 
W.~Da~Silva$^{8}$, 
E.~Dane'$^{18}$, 
P.~David$^{8}$, 
I.~De~Bonis$^{4}$, 
S.~De~Capua$^{21,k}$, 
M.~De~Cian$^{39}$, 
F.~De~Lorenzi$^{12}$, 
J.M.~De~Miranda$^{1}$, 
L.~De~Paula$^{2}$, 
P.~De~Simone$^{18}$, 
D.~Decamp$^{4}$, 
H.~Degaudenzi$^{38,37}$, 
M.~Deissenroth$^{11}$, 
L.~Del~Buono$^{8}$, 
C.~Deplano$^{15}$, 
O.~Deschamps$^{5}$, 
F.~Dettori$^{15,d}$, 
J.~Dickens$^{43}$, 
H.~Dijkstra$^{37}$, 
M.~Dima$^{28}$, 
S.~Donleavy$^{48}$, 
P.~Dornan$^{49}$, 
D.~Dossett$^{44}$, 
A.~Dovbnya$^{40}$, 
F.~Dupertuis$^{38}$, 
R.~Dzhelyadin$^{34}$, 
C.~Eames$^{49}$, 
S.~Easo$^{45}$, 
U.~Egede$^{49}$, 
V.~Egorychev$^{30}$, 
S.~Eidelman$^{33}$, 
D.~van~Eijk$^{23}$, 
F.~Eisele$^{11}$, 
S.~Eisenhardt$^{46}$, 
L.~Eklund$^{47}$, 
D.G.~d'Enterria$^{35,o}$, 
D.~Esperante~Pereira$^{36}$, 
L.~Est\`{e}ve$^{43}$, 
E.~Fanchini$^{20,j}$, 
C.~F\"{a}rber$^{11}$, 
G.~Fardell$^{46}$, 
C.~Farinelli$^{23}$, 
S.~Farry$^{12}$, 
V.~Fave$^{38}$, 
V.~Fernandez~Albor$^{36}$, 
M.~Ferro-Luzzi$^{37}$, 
S.~Filippov$^{32}$, 
C.~Fitzpatrick$^{46}$, 
W.~Flegel$^{37}$, 
F.~Fontanelli$^{19,i}$, 
R.~Forty$^{37}$, 
M.~Frank$^{37}$, 
C.~Frei$^{37}$, 
M.~Frosini$^{17,f}$, 
J.L.~Fungueirino~Pazos$^{36}$, 
S.~Furcas$^{20}$, 
A.~Gallas~Torreira$^{36}$, 
D.~Galli$^{14,c}$, 
M.~Gandelman$^{2}$, 
P.~Gandini$^{51}$, 
Y.~Gao$^{3}$, 
J-C.~Garnier$^{37}$, 
J.~Garofoli$^{52}$, 
L.~Garrido$^{35}$, 
C.~Gaspar$^{37}$, 
J.~Gassner$^{39}$, 
N.~Gauvin$^{38}$, 
P.~Gavillet$^{37}$, 
M.~Gersabeck$^{37}$, 
T.~Gershon$^{44}$, 
Ph.~Ghez$^{4}$, 
V.~Gibson$^{43}$, 
V.V.~Gligorov$^{37}$, 
C.~G\"{o}bel$^{54}$, 
D.~Golubkov$^{30}$, 
A.~Golutvin$^{49,30,37}$, 
A.~Gomes$^{2}$, 
G.~Gong$^{3}$, 
H.~Gong$^{3}$, 
H.~Gordon$^{51}$, 
M.~Grabalosa~G\'{a}ndara$^{35}$, 
R.~Graciani~Diaz$^{35}$, 
L.A.~Granado~Cardoso$^{37}$, 
E.~Graug\'{e}s$^{35}$, 
G.~Graziani$^{17}$, 
A.~Grecu$^{28}$, 
S.~Gregson$^{43}$, 
B.~Gui$^{52}$, 
E.~Gushchin$^{32}$, 
Yu.~Guz$^{34,37}$, 
Z.~Guzik$^{27}$, 
T.~Gys$^{37}$, 
G.~Haefeli$^{38}$, 
S.C.~Haines$^{43}$, 
T.~Hampson$^{42}$, 
S.~Hansmann-Menzemer$^{11}$, 
R.~Harji$^{49}$, 
N.~Harnew$^{51}$, 
P.F.~Harrison$^{44}$, 
J.~He$^{7}$, 
K.~Hennessy$^{48}$, 
P.~Henrard$^{5}$, 
J.A.~Hernando~Morata$^{36}$, 
E.~van~Herwijnen$^{37}$, 
A.~Hicheur$^{38}$, 
E.~Hicks$^{48}$, 
H.J.~Hilke$^{37}$, 
W.~Hofmann$^{10}$, 
K.~Holubyev$^{11}$, 
P.~Hopchev$^{4}$, 
W.~Hulsbergen$^{23}$, 
P.~Hunt$^{51}$, 
T.~Huse$^{48}$, 
R.S.~Huston$^{12}$, 
D.~Hutchcroft$^{48}$, 
V.~Iakovenko$^{7,41}$, 
C.~Iglesias~Escudero$^{36}$, 
C.~Ilgner$^{9}$, 
P.~Ilten$^{12}$, 
J.~Imong$^{42}$, 
R.~Jacobsson$^{37}$, 
M.~Jahjah~Hussein$^{5}$, 
E.~Jans$^{23}$, 
F.~Jansen$^{23}$, 
P.~Jaton$^{38}$, 
B.~Jean-Marie$^{7}$, 
F.~Jing$^{3}$, 
M.~John$^{51}$, 
D.~Johnson$^{51}$, 
C.R.~Jones$^{43}$, 
B.~Jost$^{37}$, 
F.~Kapusta$^{8}$, 
T.M.~Karbach$^{9}$, 
A.~Kashchuk$^{29}$, 
J.~Keaveney$^{12}$, 
U.~Kerzel$^{37}$, 
T.~Ketel$^{24}$, 
A.~Keune$^{38}$, 
B.~Khanji$^{6}$, 
Y.M.~Kim$^{46}$, 
M.~Knecht$^{38}$, 
S.~Koblitz$^{37}$, 
A.~Konoplyannikov$^{30}$, 
P.~Koppenburg$^{23}$, 
M.~Korolev$^{31}$, 
A.~Kozlinskiy$^{23}$, 
L.~Kravchuk$^{32}$, 
G.~Krocker$^{11}$, 
P.~Krokovny$^{11}$, 
F.~Kruse$^{9}$, 
K.~Kruzelecki$^{37}$, 
M.~Kucharczyk$^{25}$, 
S.~Kukulak$^{25}$, 
R.~Kumar$^{14,37}$, 
T.~Kvaratskheliya$^{30}$, 
V.N.~La~Thi$^{38}$, 
D.~Lacarrere$^{37}$, 
G.~Lafferty$^{50}$, 
A.~Lai$^{15}$, 
R.W.~Lambert$^{37}$, 
G.~Lanfranchi$^{18}$, 
C.~Langenbruch$^{11}$, 
T.~Latham$^{44}$, 
R.~Le~Gac$^{6}$, 
J.~van~Leerdam$^{23}$, 
J.-P.~Lees$^{4}$, 
R.~Lef\`{e}vre$^{5}$, 
A.~Leflat$^{31,37}$, 
J.~Lefran\c{c}ois$^{7}$, 
F.~Lehner$^{39}$, 
O.~Leroy$^{6}$, 
T.~Lesiak$^{25}$, 
L.~Li$^{3}$, 
Y.Y.~Li$^{43}$, 
L.~Li~Gioi$^{5}$, 
J.~Libby$^{51}$, 
M.~Lieng$^{9}$, 
M.~Liles$^{48}$, 
R.~Lindner$^{37}$, 
C.~Linn$^{11}$, 
B.~Liu$^{3}$, 
G.~Liu$^{37}$, 
S.~L\"{o}chner$^{10}$, 
J.H.~Lopes$^{2}$, 
E.~Lopez~Asamar$^{35}$, 
N.~Lopez-March$^{38}$, 
J.~Luisier$^{38}$, 
B.~M'charek$^{24}$, 
F.~Machefert$^{7}$, 
I.V.~Machikhiliyan$^{4,30}$, 
F.~Maciuc$^{10}$, 
O.~Maev$^{29}$, 
J.~Magnin$^{1}$, 
A.~Maier$^{37}$, 
S.~Malde$^{51}$, 
R.M.D.~Mamunur$^{37}$, 
G.~Manca$^{15,d,37}$, 
G.~Mancinelli$^{6}$, 
N.~Mangiafave$^{43}$, 
U.~Marconi$^{14}$, 
R.~M\"{a}rki$^{38}$, 
J.~Marks$^{11}$, 
G.~Martellotti$^{22}$, 
A.~Martens$^{7}$, 
L.~Martin$^{51}$, 
A.~Martin~Sanchez$^{7}$, 
D.~Martinez~Santos$^{37}$, 
A.~Massafferri$^{1}$, 
Z.~Mathe$^{12}$, 
C.~Matteuzzi$^{20}$, 
M.~Matveev$^{29}$, 
V.~Matveev$^{34}$, 
E.~Maurice$^{6}$, 
B.~Maynard$^{52}$, 
A.~Mazurov$^{32}$, 
G.~McGregor$^{50}$, 
R.~McNulty$^{12}$, 
C.~Mclean$^{46}$, 
M.~Meissner$^{11}$, 
M.~Merk$^{23}$, 
J.~Merkel$^{9}$, 
M.~Merkin$^{31}$, 
R.~Messi$^{21,k}$, 
S.~Miglioranzi$^{37}$, 
D.A.~Milanes$^{13}$, 
M.-N.~Minard$^{4}$, 
S.~Monteil$^{5}$, 
D.~Moran$^{12}$, 
P.~Morawski$^{25}$, 
J.V.~Morris$^{45}$, 
J.~Moscicki$^{37}$, 
R.~Mountain$^{52}$, 
I.~Mous$^{23}$, 
F.~Muheim$^{46}$, 
K.~M\"{u}ller$^{39}$, 
R.~Muresan$^{38}$, 
F.~Murtas$^{18}$, 
B.~Muryn$^{26}$, 
M.~Musy$^{35}$, 
J.~Mylroie-Smith$^{48}$, 
P.~Naik$^{42}$, 
T.~Nakada$^{38}$, 
R.~Nandakumar$^{45}$, 
J.~Nardulli$^{45}$, 
A.~Nawrot$^{27}$, 
M.~Nedos$^{9}$, 
M.~Needham$^{46}$, 
N.~Neufeld$^{37}$, 
P.~Neustroev$^{29}$, 
M.~Nicol$^{7}$, 
S.~Nies$^{9}$, 
V.~Niess$^{5}$, 
N.~Nikitin$^{31}$, 
A.~Oblakowska-Mucha$^{26}$, 
V.~Obraztsov$^{34}$, 
S.~Oggero$^{23}$, 
O.~Okhrimenko$^{41}$, 
R.~Oldeman$^{15,d}$, 
M.~Orlandea$^{28}$, 
A.~Ostankov$^{34}$, 
B.~Pal$^{52}$, 
J.~Palacios$^{39}$, 
M.~Palutan$^{18}$, 
J.~Panman$^{37}$, 
A.~Papanestis$^{45}$, 
M.~Pappagallo$^{13,b}$, 
C.~Parkes$^{47,37}$, 
C.J.~Parkinson$^{49}$, 
G.~Passaleva$^{17}$, 
G.D.~Patel$^{48}$, 
M.~Patel$^{49}$, 
S.K.~Paterson$^{49,37}$, 
G.N.~Patrick$^{45}$, 
C.~Patrignani$^{19,i}$, 
E.~Pauna$^{28}$, 
C.~Pauna~(Chiojdeanu)$^{28}$, 
C.~Pavel~(Nicorescu)$^{28}$, 
A.~Pazos~Alvarez$^{36}$, 
A.~Pellegrino$^{23}$, 
G.~Penso$^{22,l}$, 
M.~Pepe~Altarelli$^{37}$, 
S.~Perazzini$^{14,c}$, 
D.L.~Perego$^{20,j}$, 
E.~Perez~Trigo$^{36}$, 
A.~P\'{e}rez-Calero~Yzquierdo$^{35}$, 
P.~Perret$^{5}$, 
G.~Pessina$^{20}$, 
A.~Petrella$^{16,e,37}$, 
A.~Petrolini$^{19,i}$, 
B.~Pie~Valls$^{35}$, 
B.~Pietrzyk$^{4}$, 
D.~Pinci$^{22}$, 
R.~Plackett$^{47}$, 
S.~Playfer$^{46}$, 
M.~Plo~Casasus$^{36}$, 
G.~Polok$^{25}$, 
A.~Poluektov$^{44,33}$, 
E.~Polycarpo$^{2}$, 
D.~Popov$^{10}$, 
B.~Popovici$^{28}$, 
C.~Potterat$^{38}$, 
A.~Powell$^{51}$, 
S.~Pozzi$^{16,e}$, 
T.~du~Pree$^{23}$, 
V.~Pugatch$^{41}$, 
A.~Puig~Navarro$^{35}$, 
W.~Qian$^{3}$, 
J.H.~Rademacker$^{42}$, 
B.~Rakotomiaramanana$^{38}$, 
I.~Raniuk$^{40}$, 
G.~Raven$^{24}$, 
S.~Redford$^{51}$, 
W.~Reece$^{49}$, 
A.C.~dos~Reis$^{1}$, 
S.~Ricciardi$^{45}$, 
K.~Rinnert$^{48}$, 
D.A.~Roa~Romero$^{5}$, 
P.~Robbe$^{7,37}$, 
E.~Rodrigues$^{47}$, 
F.~Rodrigues$^{2}$, 
C.~Rodriguez~Cobo$^{36}$, 
P.~Rodriguez~Perez$^{36}$, 
G.J.~Rogers$^{43}$, 
V.~Romanovsky$^{34}$, 
J.~Rouvinet$^{38}$, 
T.~Ruf$^{37}$, 
H.~Ruiz$^{35}$, 
V.~Rusinov$^{30}$, 
G.~Sabatino$^{21,k}$, 
J.J.~Saborido~Silva$^{36}$, 
N.~Sagidova$^{29}$, 
P.~Sail$^{47}$, 
B.~Saitta$^{15,d}$, 
C.~Salzmann$^{39}$, 
A.~Sambade~Varela$^{37}$, 
M.~Sannino$^{19,i}$, 
R.~Santacesaria$^{22}$, 
R.~Santinelli$^{37}$, 
E.~Santovetti$^{21,k}$, 
M.~Sapunov$^{6}$, 
A.~Saputi$^{18}$, 
A.~Sarti$^{18}$, 
C.~Satriano$^{22,m}$, 
A.~Satta$^{21}$, 
M.~Savrie$^{16,e}$, 
D.~Savrina$^{30}$, 
P.~Schaack$^{49}$, 
M.~Schiller$^{11}$, 
S.~Schleich$^{9}$, 
M.~Schmelling$^{10}$, 
B.~Schmidt$^{37}$, 
O.~Schneider$^{38}$, 
T.~Schneider$^{37}$, 
A.~Schopper$^{37}$, 
M.-H.~Schune$^{7}$, 
R.~Schwemmer$^{37}$, 
A.~Sciubba$^{18,l}$, 
M.~Seco$^{36}$, 
A.~Semennikov$^{30}$, 
K.~Senderowska$^{26}$, 
N.~Serra$^{23}$, 
J.~Serrano$^{6}$, 
B.~Shao$^{3}$, 
M.~Shapkin$^{34}$, 
I.~Shapoval$^{40,37}$, 
P.~Shatalov$^{30}$, 
Y.~Shcheglov$^{29}$, 
T.~Shears$^{48}$, 
L.~Shekhtman$^{33}$, 
O.~Shevchenko$^{40}$, 
V.~Shevchenko$^{30}$, 
A.~Shires$^{49}$, 
E.~Simioni$^{24}$, 
H.P.~Skottowe$^{43}$, 
T.~Skwarnicki$^{52}$, 
N.~Smale$^{10}$, 
A.~Smith$^{37}$, 
A.C.~Smith$^{37}$, 
K.~Sobczak$^{5}$, 
F.J.P.~Soler$^{47}$, 
A.~Solomin$^{42}$, 
P.~Somogy$^{37}$, 
F.~Soomro$^{49}$, 
B.~Souza~De~Paula$^{2}$, 
B.~Spaan$^{9}$, 
A.~Sparkes$^{46}$, 
E.~Spiridenkov$^{29}$, 
P.~Spradlin$^{51}$, 
A.~Srednicki$^{27}$, 
F.~Stagni$^{37}$, 
S.~Steiner$^{39}$, 
O.~Steinkamp$^{39}$, 
O.~Stenyakin$^{34}$, 
S.~Stoica$^{28}$, 
S.~Stone$^{52}$, 
B.~Storaci$^{23}$, 
U.~Straumann$^{39}$, 
N.~Styles$^{46}$, 
M.~Szczekowski$^{27}$, 
P.~Szczypka$^{38}$, 
T~Szumlak$^{26}$, 
S.~T'Jampens$^{4}$, 
V.~Talanov$^{34}$, 
E.~Tarkovskiy$^{30}$, 
E.~Teodorescu$^{28}$, 
H.~Terrier$^{23}$, 
F.~Teubert$^{37}$, 
C.~Thomas$^{51,45}$, 
E.~Thomas$^{37}$, 
J.~van~Tilburg$^{39}$, 
V.~Tisserand$^{4}$, 
M.~Tobin$^{39}$, 
S.~Topp-Joergensen$^{51}$, 
M.T.~Tran$^{38}$, 
S.~Traynor$^{12}$, 
U.~Trunk$^{10}$, 
A.~Tsaregorodtsev$^{6}$, 
N.~Tuning$^{23}$, 
A.~Ukleja$^{27}$, 
P.~Urquijo$^{52}$, 
U.~Uwer$^{11}$, 
V.~Vagnoni$^{14}$, 
G.~Valenti$^{14}$, 
R.~Vazquez~Gomez$^{35}$, 
P.~Vazquez~Regueiro$^{36}$, 
S.~Vecchi$^{16}$, 
J.J.~Velthuis$^{42}$, 
M.~Veltri$^{17,g}$, 
K.~Vervink$^{37}$, 
B.~Viaud$^{7}$, 
I.~Videau$^{7}$, 
X.~Vilasis-Cardona$^{35,n}$, 
J.~Visniakov$^{36}$, 
A.~Vollhardt$^{39}$, 
D.~Voong$^{42}$, 
A.~Vorobyev$^{29}$, 
An.~Vorobyev$^{29}$, 
H.~Voss$^{10}$, 
K.~Wacker$^{9}$, 
S.~Wandernoth$^{11}$, 
J.~Wang$^{52}$, 
D.R.~Ward$^{43}$, 
A.D.~Webber$^{50}$, 
D.~Websdale$^{49}$, 
M.~Whitehead$^{44}$, 
D.~Wiedner$^{11}$, 
L.~Wiggers$^{23}$, 
G.~Wilkinson$^{51}$, 
M.P.~Williams$^{44,45}$, 
M.~Williams$^{49}$, 
F.F.~Wilson$^{45}$, 
J.~Wishahi$^{9}$, 
M.~Witek$^{25}$, 
W.~Witzeling$^{37}$, 
S.A.~Wotton$^{43}$, 
K.~Wyllie$^{37}$, 
Y.~Xie$^{46}$, 
F.~Xing$^{51}$, 
Z.~Yang$^{3}$, 
G.~Ybeles~Smit$^{23}$, 
R.~Young$^{46}$, 
O.~Yushchenko$^{34}$, 
M.~Zavertyaev$^{10,a}$, 
M.~Zeng$^{3}$, 
L.~Zhang$^{52}$, 
W.C.~Zhang$^{12}$, 
Y.~Zhang$^{3}$, 
A.~Zhelezov$^{11}$, 
L.~Zhong$^{3}$, 
E.~Zverev$^{31}$.\bigskip\newline{\it\footnotesize
$ ^{1}$Centro Brasileiro de Pesquisas F\'{i}sicas (CBPF), Rio de Janeiro, Brazil\\
$ ^{2}$Universidade Federal do Rio de Janeiro (UFRJ), Rio de Janeiro, Brazil\\
$ ^{3}$Center for High Energy Physics, Tsinghua University, Beijing, China\\
$ ^{4}$LAPP, Universit\'{e} de Savoie, CNRS/IN2P3, Annecy-Le-Vieux, France\\
$ ^{5}$Clermont Universit\'{e}, Universit\'{e} Blaise Pascal, CNRS/IN2P3, LPC, Clermont-Ferrand, France\\
$ ^{6}$CPPM, Aix-Marseille Universit\'{e}, CNRS/IN2P3, Marseille, France\\
$ ^{7}$LAL, Universit\'{e} Paris-Sud, CNRS/IN2P3, Orsay, France\\
$ ^{8}$LPNHE, Universit\'{e} Pierre et Marie Curie, Universit\'{e} Paris Diderot, CNRS/IN2P3, Paris, France\\
$ ^{9}$Fakult\"{a}t Physik, Technische Universit\"{a}t Dortmund, Dortmund, Germany\\
$ ^{10}$Max-Planck-Institut f\"{u}r Kernphysik (MPIK), Heidelberg, Germany\\
$ ^{11}$Physikalisches Institut, Ruprecht-Karls-Universit\"{a}t Heidelberg, Heidelberg, Germany\\
$ ^{12}$School of Physics, University College Dublin, Dublin, Ireland\\
$ ^{13}$Sezione INFN di Bari, Bari, Italy\\
$ ^{14}$Sezione INFN di Bologna, Bologna, Italy\\
$ ^{15}$Sezione INFN di Cagliari, Cagliari, Italy\\
$ ^{16}$Sezione INFN di Ferrara, Ferrara, Italy\\
$ ^{17}$Sezione INFN di Firenze, Firenze, Italy\\
$ ^{18}$Laboratori Nazionali dell'INFN di Frascati, Frascati, Italy\\
$ ^{19}$Sezione INFN di Genova, Genova, Italy\\
$ ^{20}$Sezione INFN di Milano Bicocca, Milano, Italy\\
$ ^{21}$Sezione INFN di Roma Tor Vergata, Roma, Italy\\
$ ^{22}$Sezione INFN di Roma Sapienza, Roma, Italy\\
$ ^{23}$Nikhef National Institute for Subatomic Physics, Amsterdam, Netherlands\\
$ ^{24}$Nikhef National Institute for Subatomic Physics and Vrije Universiteit, Amsterdam, Netherlands\\
$ ^{25}$Henryk Niewodniczanski Institute of Nuclear Physics  Polish Academy of Sciences, Cracow, Poland\\
$ ^{26}$Faculty of Physics \& Applied Computer Science, Cracow, Poland\\
$ ^{27}$Soltan Institute for Nuclear Studies, Warsaw, Poland\\
$ ^{28}$Horia Hulubei National Institute of Physics and Nuclear Engineering, Bucharest-Magurele, Romania\\
$ ^{29}$Petersburg Nuclear Physics Institute (PNPI), Gatchina, Russia\\
$ ^{30}$Institute of Theoretical and Experimental Physics (ITEP), Moscow, Russia\\
$ ^{31}$Institute of Nuclear Physics, Moscow State University (SINP MSU), Moscow, Russia\\
$ ^{32}$Institute for Nuclear Research of the Russian Academy of Sciences (INR RAN), Moscow, Russia\\
$ ^{33}$Budker Institute of Nuclear Physics (BINP), Novosibirsk, Russia\\
$ ^{34}$Institute for High Energy Physics(IHEP), Protvino, Russia\\
$ ^{35}$Universitat de Barcelona, Barcelona, Spain\\
$ ^{36}$Universidad de Santiago de Compostela, Santiago de Compostela, Spain\\
$ ^{37}$European Organization for Nuclear Research (CERN), Geneva, Switzerland\\
$ ^{38}$Ecole Polytechnique F\'{e}d\'{e}rale de Lausanne (EPFL), Lausanne, Switzerland\\
$ ^{39}$Physik-Institut, Universit\"{a}t Z\"{u}rich, Z\"{u}rich, Switzerland\\
$ ^{40}$NSC Kharkiv Institute of Physics and Technology (NSC KIPT), Kharkiv, Ukraine\\
$ ^{41}$Institute for Nuclear Research of the National Academy of Sciences (KINR), Kyiv, Ukraine\\
$ ^{42}$H.H. Wills Physics Laboratory, University of Bristol, Bristol, United Kingdom\\
$ ^{43}$Cavendish Laboratory, University of Cambridge, Cambridge, United Kingdom\\
$ ^{44}$Department of Physics, University of Warwick, Coventry, United Kingdom\\
$ ^{45}$STFC Rutherford Appleton Laboratory, Didcot, United Kingdom\\
$ ^{46}$School of Physics and Astronomy, University of Edinburgh, Edinburgh, United Kingdom\\
$ ^{47}$School of Physics and Astronomy, University of Glasgow, Glasgow, United Kingdom\\
$ ^{48}$Oliver Lodge Laboratory, University of Liverpool, Liverpool, United Kingdom\\
$ ^{49}$Imperial College London, London, United Kingdom\\
$ ^{50}$School of Physics and Astronomy, University of Manchester, Manchester, United Kingdom\\
$ ^{51}$Department of Physics, University of Oxford, Oxford, United Kingdom\\
$ ^{52}$Syracuse University, Syracuse, NY, United States of America\\
$ ^{53}$CC-IN2P3, CNRS/IN2P3, Lyon-Villeurbanne, France, associated member\\
$ ^{54}$Pontif\'{i}cia Universidade Cat\'{o}lica do Rio de Janeiro (PUC-Rio), Rio de Janeiro, Brazil, associated to $^2 $\\
\bigskip
$ ^{a}$P.N. Lebedev Physical Institute, Russian Academy of Science (LPI RAS), Moskow, Russia\\
$ ^{b}$Universit\`{a} di Bari, Bari, Italy\\
$ ^{c}$Universit\`{a} di Bologna, Bologna, Italy\\
$ ^{d}$Universit\`{a} di Cagliari, Cagliari, Italy\\
$ ^{e}$Universit\`{a} di Ferrara, Ferrara, Italy\\
$ ^{f}$Universit\`{a} di Firenze, Firenze, Italy\\
$ ^{g}$Universit\`{a} di Urbino, Urbino, Italy\\
$ ^{h}$Universit\`{a} di Modena e Reggio Emilia, Modena, Italy\\
$ ^{i}$Universit\`{a} di Genova, Genova, Italy\\
$ ^{j}$Universit\`{a} di Milano Bicocca, Milano, Italy\\
$ ^{k}$Universit\`{a} di Roma Tor Vergata, Roma, Italy\\
$ ^{l}$Universit\`{a} di Roma La Sapienza, Roma, Italy\\
$ ^{m}$Universit\`{a} della Basilicata, Potenza, Italy\\
$ ^{n}$LIFAELS, La Salle, Universitat Ramon Llull, Barcelona, Spain\\
$ ^{o}$Instituci\'{o} Catalana de Recerca i Estudis Avan\c{c}ats (ICREA), Barcelona, Spain\\
}
\end{flushleft}

\cleardoublepage
\setcounter{page}{1}
\pagenumbering{arabic}

\newpage
\section{Introduction}
\noindent Much less is known experimentally about semileptonic $\Bs$ decays, than for the lighter $B$ mesons. In the case of the $\Bs$ when the $b\to c$ transition results in a single charm hadron this can be a $D_s^+$, a $D_s^{*+}$ or another excited $c\overline{s}$ state. The relative proportion of these final states provides essential information on the structure of these semileptonic decays, and can be compared with QCD-based theoretical models. In this Letter we present a search for $\Bs$ semileptonic decays, that might occur via an excited  $c\overline{s}$ meson that disintegrates into final states containing $D^0K^+$.  One such state is the $D_{s1}^+$, thought to be $J^P=1^+$, that decays into $D^*K$, and another is the $D_{s2}^{*+}$, a possible $2^+$ state that has been observed to decay directly into $DK$ \cite{PDG}.

The LHCb detector \cite{LHCb-det} is a forward spectrometer constructed primarily to measure $CP$-violating and rare decays of hadrons containing $b$ and $c$ quarks.
The detector elements are placed along the beam line of the LHC starting with the Vertex Locator (VELO), a silicon strip device that surrounds the proton-proton interaction region and is positioned 8 mm from the beam during collisions. The VELO precisely determines the locations of primary $pp$ interaction vertices, the locations of decays of long lived hadrons, and contributes to the measurement of track momenta.  Other detectors used to measure track momenta comprise a large area silicon strip detector (TT)  located before a 3.7 Tm dipole magnet, and a combination of
silicon strip detectors (IT) and straw drift chambers (OT) placed afterward. Two Ring Imaging Cherenkov (RICH) detectors are used to identify charged hadrons. Further downstream an Electromagnetic Calorimeter (ECAL) is used for photon detection and electron identification, followed by a Hadron Calorimeter (HCAL), and  a system consisting of alternating layers of iron and chambers (MWPC and triple-GEM) that distinguishes muons from hadrons (MUON). The ECAL, MUON, and HCAL provide the capability of first-level hardware triggering.

	In this analysis we use a data sample of approximately 20 pb$^{-1}$  collected from 7 TeV centre-of-mass energy $pp$ collisions at the LHC during 2010. For the first 3 pb$^{-1}$ of these data a
trigger was used that requires a single muon without any requirement that it misses the primary vertex, a trigger which was not available for the remainder of the data taking. This sample is well suited to determine the number of semileptonic $\Bs$ decays, that we take as the sum of $D_s^+ X\munu$, $D^0K^+X\munu$ and $D^+K^0 X\munu$ decays, ignoring the small $\approx$1\% contribution from charmless $\Bs$ decays. The entire 20  pb$^{-1}$  sample, however, is useful
for establishing signal significance, resonance parameter determination, and the ratio of numbers of events in the $D^0K^+$ states.

\section{Selection criteria}
In both data samples backgrounds increase markedly with increasing track numbers. Thus, events are
accepted only if the number of reconstructed tracks using the VELO and
either the IT or OT is less than 100. Tracks were accepted based on similar criteria to those described in Ref.~\cite{LHCb-det}. This results in only a 5.6\% loss of
signal in the 3 pb$^{-1}$, and a larger 9.4\% loss over the entire 20  pb$^{-1}$ sample.

In this analysis we select a charm hadron that forms a vertex with an identified muon.  We consider two cases: (i)  $D_s^+\to K^+K^-\pi^+$, that has a branching fraction of (5.50$\pm$0.27)\% \cite{PDG} -- these are used to normalize the $\Bs$ yield; (ii) $D^0\to K^-\pi^+$ decays with a branching fraction of  (3.89$\pm$0.05)\% \cite{PDG} -- these are combined with an additional $K^+$  that forms a  vertex with the $D^0$ and the $\mu^-$ in order to search for $\Bs$ semileptonic decays that might occur via an excited  $c\overline{s}$ meson that decays into $D^0K^+$.  In this Letter the mention of a specific final state will refer also to its charge-conjugate state. The selection techniques are similar to those used in a previous analysis \cite{1stpaper}.
Most charm hadrons  are produced directly via $pp\to c\overline{c} X$ interactions at the LHC, where $X$ indicates the sum over all other possible final state particles. We denote these particular charm reactions as ``Prompt".  Charm is also produced in $pp\to b\overline{b} X$ collisions where the $b$-flavoured hadron decays into charm. These are called charm from $b$ hadrons or ``Dfb" for short. Muon candidates are selected using their penetration through the iron of the muon system. The candidates used in the analysis of the first 3 pb$^{-1}$ sample must be those that triggered the event and have momentum transverse to the beam direction, $p_{\rm T}$, greater than 1200 MeV (we use units with $c$=1).

The selection criteria for $D_s^+$ and $D^0$ mesons include identifying kaon and pion candidates using the RICH system. Cherenkov photon angles with respect to the track direction are examined and a likelihood formed for each particle hypothesis \cite{LHCb-det}.  We also require that the $p_{\rm T}$ of the kaons and pion be greater than 300 MeV, and that their scalar sum be greater than 2100 MeV ($D_s^+)$ or greater than 1400 MeV ($D^0$).
Since charm mesons travel before decaying, the kaon and pion tracks when followed backwards will most often not point to the  primary vertex. The impact parameter (IP) is the minimum distance of approach of the track with respect to the primary vertex. We require that the $\chi^2$ formed by using the hypothesis that the IP is equal to zero, $\chi^2_{\rm IP}$, be $>9$ for each track. The kaon and pion candidate tracks must also be consistent with coming from a common origin, the charm decay vertex,  with vertex fit $\chi^2$ per number of degrees of freedom (ndof) $<6$.
This charm candidate's decay vertex must be detached from the closest primary interaction point. To implement this flight distance significance test we form a variable, $\chi^2_{\rm FS}$, based on the hypothesis that the flight distance between the primary and charm vertices is zero, and require $\chi^2_{\rm FS}>100$. 

Partial $\Bs$ candidates formed from $D^+_s$ muon candidates must form a  vertex with $\chi^2$/ndof $<6$,  and point at the primary vertex: the cosine of the angle of the $b$ pseudo-direction formed from the $D_s^+$ and muon vector momentum sum with respect to the line between the $D_s^+\mu^-$ vertex
and the primary vertex ($\cos\delta$) must be  $>$ 0.999. They must also have an invariant mass in the range 3.10~GeV$<m(D_s^+\mu^-)<5.10$ GeV.
All of these requirements were decided upon by comparing the sidebands of the invariant mass distributions, representative of the background, with signal Monte Carlo simulation using PYTHIA 6.4 \cite{Pythia} event generation, and the GEANT4 \cite{GEANT4} based LHCb detector simulation.

\begin{figure}[hbt]
\centering
\includegraphics[width=6.in]{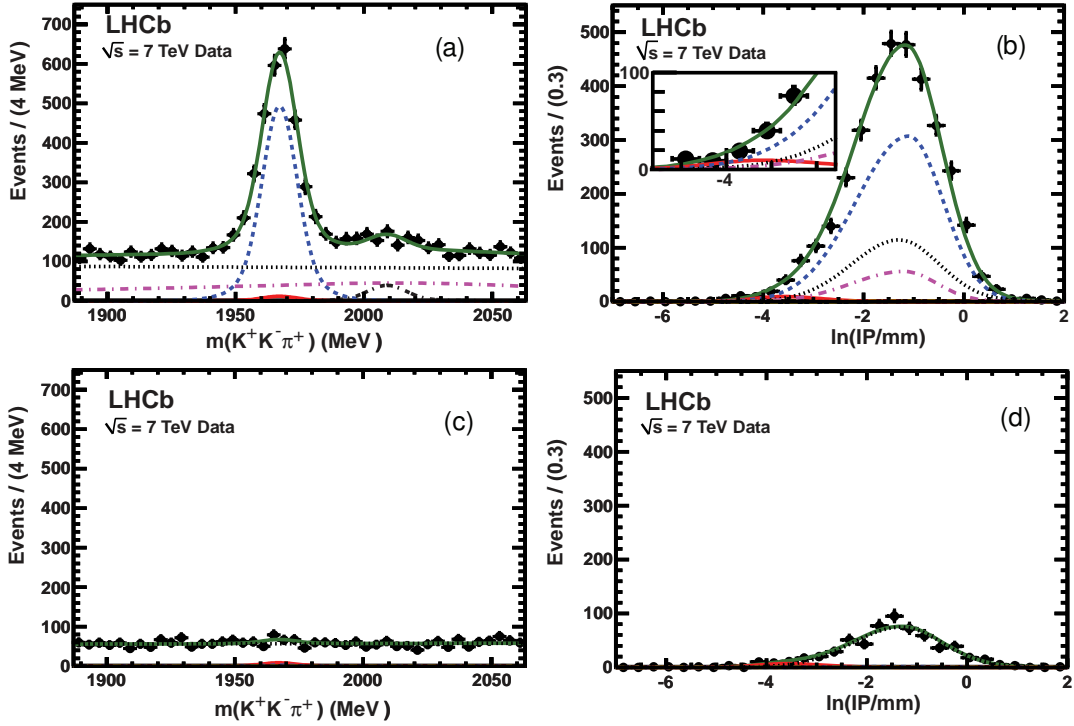}
\caption{The invariant $K^+K^-\pi^+$ mass spectra for events associated with a muon for the 3 pb$^{-1}$ sample in the pseudorapidity interval $2<\eta<6$ for RS combinations (a) and WS combinations (c). Also shown is the natural logarithm of the IP distributions of the $D_s^+$ candidates for (b) RS and (d) WS $D^+_s$ muon candidate combinations. The labelling of the curves is the same on all four sub-figures. In descending order in (a):  green-solid curve shows the total, the blue-dashed curve the Dfb signal, the 
black-dotted curve the sideband background, the purple-dot-dashed the misinterpreted $\Lambda_c^+\to pK^-\pi^+$ contribution, the black dash-dash-dot curve the $D^{*+}\to \pi^+D^0\to K^+K^-\pi^+$ contribution, and
the barely visible red-solid curves the Prompt yield. The Dfb signal, the $\Lambda_c^+$ reflection and $D^{*+}$ signal are too small to be seen in the WS distributions. The insert in (b) shows an expanded view of the region populated by Prompt charm production.
 } \label{Ds-overall-Lc}
\end{figure}

The analysis for the $D_s^+ X\munu$ mode follows the same procedure as our previous $D^0 X \munu$ study \cite{1stpaper}, and uses the 3 pb$^{-1}$ sample.
The $K^+K^-\pi^+$ mass spectra for both the right-sign  (RS $K^+K^-\pi^+$ + $\mu^-$) and wrong-sign (WS $K^+K^-\pi^+$ + $\mu^+$) candidates, as well as the ln(IP/mm) distributions for events with mass combinations within $\pm$20 MeV of the $D_s^+$ mass are shown in Fig.~\ref{Ds-overall-Lc} for the pseudorapidity interval $2<\eta<6$. Here IP refers to the impact parameter of the $D_s^+$ candidate with respect to the primary vertex in units of mm.  For both the RS and WS cases, we perform unbinned extended maximum likelihood fits to the two-dimensional distributions in $K^+K^-\pi^+$ invariant mass and ln(IP/mm), over a region extending  from 80 MeV below the $D_s^+$ mass peak to 96 MeV above.  This
fitting procedure allows us to determine directly the background shape from false combinations under the $D_s^+$ signal mass peak. The parameters of the Prompt IP distribution are found by examining directly produced charm \cite{1stpaper}. The Monte Carlo simulated shape is used for the Dfb component.  The fit separates
contributions from Dfb, Prompt, and false combinations. The Prompt contribution is small. Background components for $D^{*+}\to \pi^+D^0\to \pi^+K^+K^-$ and the reflection from  $\Lambda_c^+\to pK^-\pi^+$ decay, where either a proton or a pion is wrongly identified as a kaon by the particle identification system, are also included. The shape of the $D^{*+}$ background is constrained to be equal to that of the $D_s^+\to K^+K^-\pi^+$ signal peak and the yield is allowed to float, while the shape of the $\Lambda_c^+$ reflection is determined from Monte Carlo and the yield is 
allowed to float within the uncertainty of our expectation.

To evaluate more carefully the $D_s^+$ yield 
the fits are performed in $\eta$ bins and the detection efficiency in each bin is determined separately so as to remove uncertainty from differences in the $\eta$ dependent production observed in data compared to the Monte Carlo simulation.
This procedure yields  2233$\pm$60 RS Dfb events in the $D_s^+ X \munu$ channel in the $b$ pseudorapidity range $2<\eta<6$, uncorrected for efficiency;  the average detection efficiency is (1.07$\pm$0.03)\%. This yield is then reduced by 5.1\% for additional correlated $b$ decay backgrounds as determined by simulation.

\section{\boldmath Measurement of $D^0 K^+ X\munu$}

Semileptonic decays of $\Bs$ mesons usually result in a $D_s^+$ meson in the final state. It is possible, however, that the semileptonic decay goes to a $c\overline{s}$ excitation,  which can decay into either $D K$ or $D^* K$ resonances, or produces non-resonant $DK$.  To search for these final states, we measure the $D^0 K^+ X \munu$ yield. 
To seek events with a $D^0$ candidate and an additional $K^+$ we require that
the $K^+$ candidate has $p_{\rm T} > 300$ MeV, be identified as such in the RICH system, has $\chi^2_{\rm IP}>9$, and that the vector sum $p_{\rm T}$  of the $D^0$ and kaon be $>1500$ MeV. The resulting partial $B$ candidate must have an invariant mass in the range 3.09~GeV$<m(D^0K^+\mu^-)<5.09$ GeV, form a  vertex ($\chi^2$/ndof $<$ 3) and point at the primary vertex ($\cos\delta>$0.999). In addition, we explicitly check that if the kaon candidate is assigned the pion mass and combined with the $D^0$, it does not form a  $D^{*+}$ candidate, by requiring the difference in masses  $m(K^-\pi^+\pi^+) - m(K^-\pi^+) - m(\pi^+)> 20$ MeV, in addition to the $\pm$20 MeV requirement around the $D^0$ mass for $m(K^-\pi^+)$.

\begin{figure}[hbt]
\centering
\includegraphics[width=5.in]{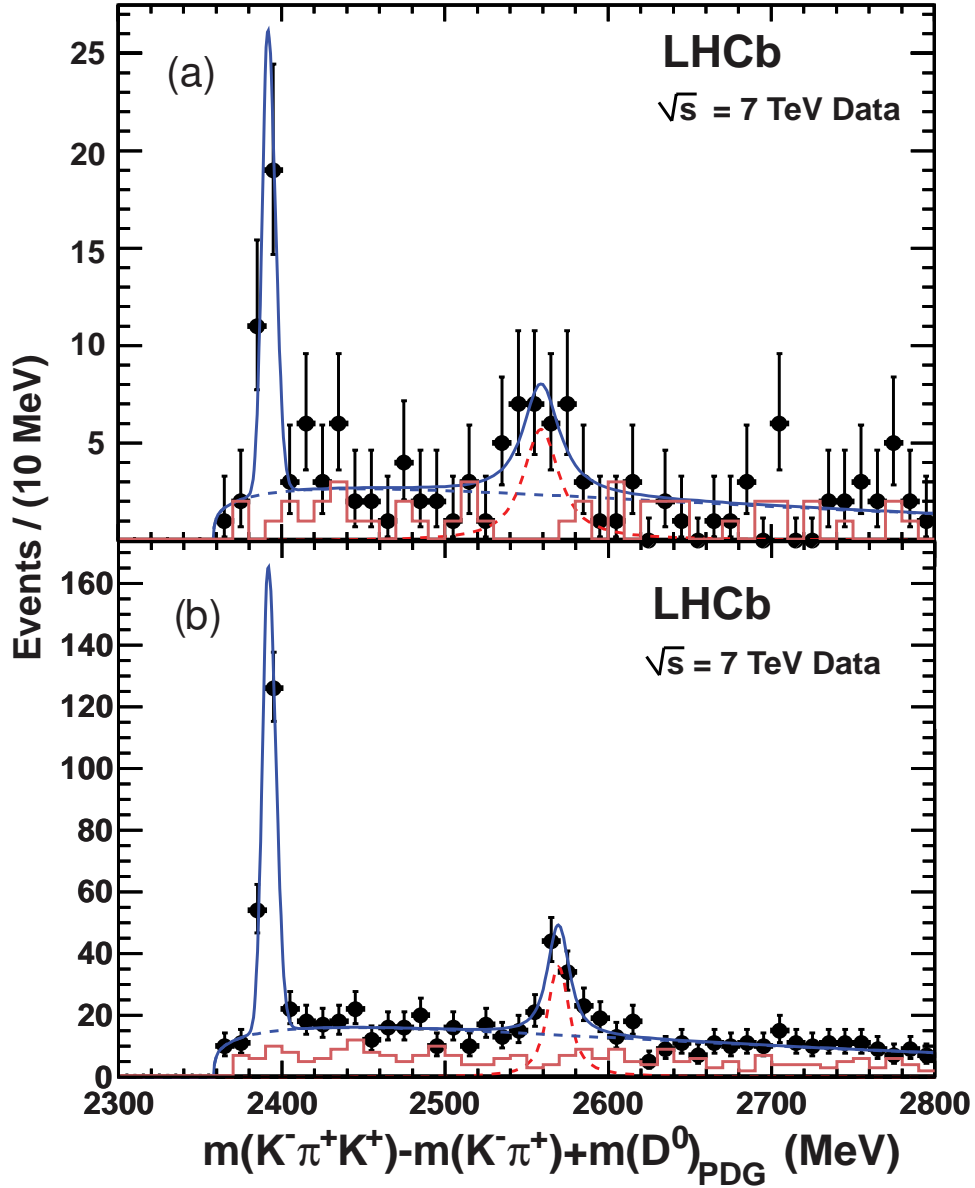}
\caption{The mass difference $m(K^-\pi^+K^+)-m(K^-\pi^+)$ added to the known $D^0$ mass for events with $K^-\pi^+$ invariant masses within $\pm$20 MeV of the $D^0$ mass (black points) in semileptonic decays. The histogram shows wrong-sign events with an additional $K^-$ instead of a $K^+$. The curves are described in the text. (a) For the 3 pb$^{-1}$ data sample and (b) for the 20 pb$^{-1}$ sample.
 } \label{m_d0k_vetop}
\end{figure}

Figure~\ref{m_d0k_vetop}(a) shows the $D^0 K^+$ invariant mass spectrum in the 3 pb$^{-1}$ sample. $D^0$ candidates are chosen from $K^-\pi^+ X\munu$ events with a $K^-\pi^+$ invariant mass within $\pm$20 MeV of the $D^0$ mass.
\afterpage{\clearpage}
A clear narrow signal near threshold is seen corresponding to the $D_{s1}(2536)^+$, but at a lower mass of 2392 MeV.
An axial-vector state cannot decay into two pseudoscalar mesons but this resonance can decay into $D^{*0}K^+$. Since we do not reconstruct the $\gamma$ or $\pi^0$ from the $D^{*0}$, the mass peak will be shifted down from its nominal value.  However, because the resonance is so close to threshold, the mass resolution will still be very good resulting in a narrow peak.
This final state was seen previously in $\Bs$ semileptonic decays by the D0 collaboration using  $D_{s1}^{+}\to D^{*+}K_S^0$ decays \cite{D0}. There also appears to be a feature near the known mass of the $D^*_{s2}(2573)^{+}$ meson. The width of this state is not well measured; the PDG quotes 20$\pm$5 MeV \cite{PDG}. Clearly there is a large excess over the wrong-sign background here evaluated using $D^0K^-$ mass combinations.

In order to ascertain the size of the putative signals above background we perform an unbinned maximum likelihood fit.
The data are fit with a threshold background function proportional to 
${\cal{M}}^pe^{-a{\cal{M}}}$, with ${\cal{M}}=m(D^0K^+)-m_0$,
where $m_0$, the threshold point, is fixed at 2358.52 MeV. The fit determines $p$ and $a$. We assume that the $\Bs\to D^0K^+X\munu$ signal above the background function is saturated by the $D_{s1}^{+}$ and $D_{s2}^{*+}$ states. For the $D_{s1}^+$ signal function we use a bifurcated Gaussian shape, whose relative widths above and below the peak are fixed from simulation.  The mass and average width are fixed to the values  2391.6 MeV and 4.0 MeV, respectively, found using the higher statistics sample discussed below, while the simulation, including the effects of the missing $D^{*0}$ decay product, predicts a mass of 2392.2$\pm$0.3 MeV. The width is essentially due to the missing $\gamma$ or $\pi^0$ from the $D^{*0}$ to $D^0$ decay. There are 24.4$\pm$5.5 $D_{s1}^+$ events.  A relativistic Breit-Wigner signal shape convolved with the experimental resolution of 3.3 MeV (r.m.s.) is used in the region of the $D_{s2}^{*+}$ where both the mass and width are allowed to float in the fit.  We find a mass value of 2559$\pm$9 MeV, a width of 24.1$\pm$9.2 MeV and 22.1$\pm$7.5 events, where all of these uncertainties are statistical only. 


To confirm the $D_{s2}^{*+}$ signal we use the full data sample of 20 pb$^{-1}$, in which we accept all events that were triggered. While this sample is useful to increase statistics it suffers from a larger number of interactions per crossing, and multiple triggers, that makes it more difficult to ascertain the total number of $\Bs$ decays. The measurement of the relative yields of $D_{s2}^{*+}$ to $D_{s1}^{+}$, however, will not be affected.  Figure~\ref{m_d0k_vetop}(b) shows the resulting $D^0K^+$ invariant mass spectrum. The difference between RS and WS events outside of the resonant peaks is consistent with background from other $b$ decays as demonstrated by Monte Carlo simulation.
We use the same fitting functions as above, but here we allow the mass and average width values of the bifurcated Gaussian to float while still fixing the ratio of widths above and below the peak from simulation. 
The fit to the $D_{s1}^+$ yields 155$\pm$15 signal events, a $D^0K^+$ mass of 2391.6$\pm$0.5 MeV, and 4.0$\pm$0.4 MeV for the width.
For the $D_{s2}^{*+}$ we again allow the mass, the width and the number of events to float in the fit.
We find a mass  of 2569.4$\pm$1.6 MeV,  a width of 12.1$\pm$4.5 MeV, and 82$\pm$17 events. These errors are purely statistical.   The previously measured mass and width values from the PDG are 2572.6$\pm$0.9 MeV and 20$\pm$5 MeV  \cite{PDG}. The probability of the background fluctuating to form the $D_{s2}^{*+}$ signal corresponds to eight standard deviations, as determined by the change in twice the natural logarithm of the likelihood of the fit without including this resonance and accounting for the change in the number of degrees of freedom.

The systematic uncertainty on the $D_{s2}^{*+}$ mass is determined from several calibration channels. For example, our measured $D^0$ mass differs from the known value by 0.2 MeV, though the known value has a 0.14 MeV error. We also see a variation on the order of 0.3 MeV by varying the fit region and background shape, where we use a linear function instead of the threshold function. Thus we take $\pm$0.5 MeV as the systematic uncertainty. We use the same method of changing the fits to find the systematic uncertainty on the width. The maximum observed change is 1.4 MeV. There is also a contribution from our uncertainty on the experimental resolution of $\pm$0.5 MeV that contributes an additional 0.7 MeV error on the width. Taking these two components in quadrature gives a width uncertainty of 1.6 MeV.

The relative branching fractions are determined from the 20 pb$^{-1}$ sample, assuming that the $D_{s1}^{+}$ decays only into $D^*K$ final states, the $D_{s2}^{*+}$ decays only into $DK$ final states, and isospin is conserved in their decays.  Note that the only observed decays $D_{s2}^{*+}$ are to $DK$ final states, while decays to $D^*K$, although possible, have not yet been seen, including the study by the D0 collaboration \cite{D0}. The $D_{s2}^{*+}/D_{s1}^+$ event ratio is computed, correcting for the lower detection efficiency for $D_{s2}^{*+}$ of (0.516$\pm$0.017)\%,  compared with the $D_{s1}^+$ efficiency of (0.598$\pm$0.025)\% as
\begin{equation}
\label{eq:Ds2BR}
\frac{{\cal{B}}(\Bs\to D_{s2}^{*+} X \munu)}{{\cal{B}}(\Bs\to D_{s1}^+ X \munu)}=0.61\pm 0.14\pm 0.05.
\end{equation}

The relative branching fraction of the $D_{s1}^+$ with respect to the total $B_s$ semileptonic rate
is measured using 24.4 $\pm$ 5.5 events in the 3 pb$^{-1}$ sample.
The number of $\Bs$ semileptonic decay events  in this sample is evaluated from the efficiency corrected sum of the $\Bs\to D_s^+ X \munu$  events and twice the efficiency corrected $\Bs\to D^0 X K^+\munu$ yield. The efficiencies are 1.07\% and 0.57\%, respectively. The doubling of the $D^0K^+X\munu$ yield accounts for the missing $D^+K^0 X\munu$ contribution, which is equal due to isospin symmetry.
A small component of $B\to D_s^+ K X\munu$ is subtracted based on a branching fraction measurement from BaBar of $(6.1\pm 1.2)\times 10^{-4}$ \cite{BtoDsK}, reducing the $D_s^+ X\munu$ yield by 3.2\%. The overall uncertainty on the $\Bs$ semileptonic yield is 6.6\%. The main contributions to this error are the uncertainty on the absolute $D_s^+$ branching ratio of 4.9\%, and the uncertainty on the amount of $D^0K^+X\munu$ events to add to the $\Bs$ yield
of 3.0\%. 
The corresponding number for the $D_{s2}^{*+}$ branching fraction is computed also using this sample and the result from Eq.~\ref{eq:Ds2BR}. Correcting for the unreconstructed $D^+K^0$ decays results in the doubling of the rates of the relative branching fractions, that we determine to be
\begin{eqnarray}
\frac{{\cal{B}}(\Bs\to D_{s2}^{*+} X \munu)}{{\cal{B}}(\Bs\to X \munu)}&=& (3.3\pm 1.0\pm 0.4 )\%\nonumber\\
\frac{{\cal{B}}(\Bs\to D_{s1}^+ X \munu)}{{\cal{B}}(\Bs\to X \munu)}&=& (5.4\pm 1.2\pm 0.5)\%,
\end{eqnarray}
where the systematic uncertainty for both includes a 5\% error on the detection efficiency,  and the above mentioned 6.6\% uncertainty on the number of $\Bs$ semileptonic decays. 
In addition there is a systematic uncertainty of 8\% on the  $D_{s2}^{*+}$  yield estimated by varying the fit region, and background shape. Our branching fraction for the relative rate of $D_{s1}^+$ decay is consistent with, but smaller than, the value of (9.8$\pm$3.0)\%  measured by D0 \cite{D0}.

\section{Conclusions}
The first observation has been made of
the rare semileptonic decay $\Bs\to D^*_{s2}(2573)^{+}  X \munu$ and its branching fraction relative to the total semileptonic $\Bs$ decay rate has been measured as
${{\cal{B}}(\Bs\to D_{s2}^{*+} X \munu)}/{(\Bs\to X \munu)}= (3.3\pm 1.0\pm 0.4 )\%$.
For  $\Bs\to D_{s1}(2536)^+ X \munu$ semileptonic decays the ratio is
${{\cal{B}}(\Bs\to D_{s1}^+ X \munu)}/{\cal{B}}(\Bs\to X \munu)= (5.4\pm 1.2\pm 0.4)\%$, where in both cases the first uncertainty is statistical and the second is systematic.  We have assumed that the $D_{s1}^{+}$ decays only into $D^*K$ final states, the $D_{s2}^{*+}$ decays only into $DK$ final states, and isospin is conserved in their decays. 
These values were predicted in the ISGW2 model as 3.2\% and 5.7\%, for $D_{s2}^{*+}$ and $D_{s1}^{+}$, respectively,  in good agreement with our observations \cite{ISGW2}.  Another set of predictions based on the quark model are 1.8\% and 2\%, respectively \cite{Mayorga}. The mass of the $D_{s2}^{*+}$ is measured to be 2569.4$\pm$1.6$\pm$0.5 MeV, and the width as 12.1$\pm$4.5$\pm$1.6 MeV,  in agreement with previous observations.

\section*{Acknowledgments}
We express our gratitude to our colleagues in the CERN accelerator departments for the excellent performance of the LHC.
We thank the technical and administrative staff at CERN and at the LHCb institutes, and acknowledge support from the National Agencies: CAPES, CNPq, FAPERJ and FINEP (Brazil); CERN; NSFC (China); CNRS/IN2P3 (France); BMBF, DFG, HGF and MPG (Germany); SFI (Ireland); INFN (Italy); FOM and NWO (Netherlands); SCSR (Poland); ANCS (Romania); MinES of Russia and Rosatom (Russia); MICINN, XUNGAL and GENCAT (Spain); SNSF and SER (Switzerland); NAS Ukraine (Ukraine); STFC (United Kingdom); NSF (USA). We also acknowledge the support received from the ERC under FP7 and the R\'egion Auvergne.

\end{document}